\documentclass[journal=nalefd,manuscript=letter]{achemso}
\usepackage[version=3]{mhchem} 
\usepackage[T1]{fontenc} 

\usepackage{amsmath}
\usepackage{upgreek}
\usepackage{epstopdf}

\author{Evgeny M. Alexeev}
\email{e.alexeev@sheffield.ac.uk}
\author{Alessandro Catanzaro}
\author{Oleksandr V. Skrypka}
\affiliation[University of Sheffield]
{Department of Physics and Astronomy, University of Sheffield, Sheffield S3 7RH, UK}
\author{Pramoda K. Nayak}
\author{Seongjoon Ahn}
\affiliation[UNIST]
{Department of Energy Engineering and Department of Chemistry, Ulsan National Institute of Science and Technology (UNIST), 50 UNIST-gil, Ulsan 44919, Republic of Korea}
\author{Sangyeon Pak}
\author{Juwon Lee}
\author{Jung Inn Sohn}
\affiliation[University of Oxford]
{Department of Engineering Science, University of Oxford, Oxford OX1 3PJ, UK}
\author{Kostya S. Novoselov}
\affiliation[University of Manchester]
{School of Physics and Astronomy, University of Manchester, Oxford Road, Manchester M13 9PL, UK}
\author{Hyeon Suk Shin}
\affiliation[UNIST]
{Department of Energy Engineering and Department of Chemistry, Ulsan National Institute of Science and Technology (UNIST), 50 UNIST-gil, Ulsan 44919, Republic of Korea}
\author{Alexander I. Tartakovskii}
\email{a.tartakovskii@sheffield.ac.uk}
\affiliation[University of Sheffield]
{Department of Physics and Astronomy, University of Sheffield, Sheffield S3 7RH, UK}

\title{Imaging of interlayer coupling in van der Waals heterostructures using a bright-field optical microscope}
\keywords{van der Waals heterostructures, transition metal dichalcogenides, 2D materials, interlayer coupling, photoluminescence, annealing, fluorescence imaging}
\begin{document}

\begin{tocentry}
\includegraphics{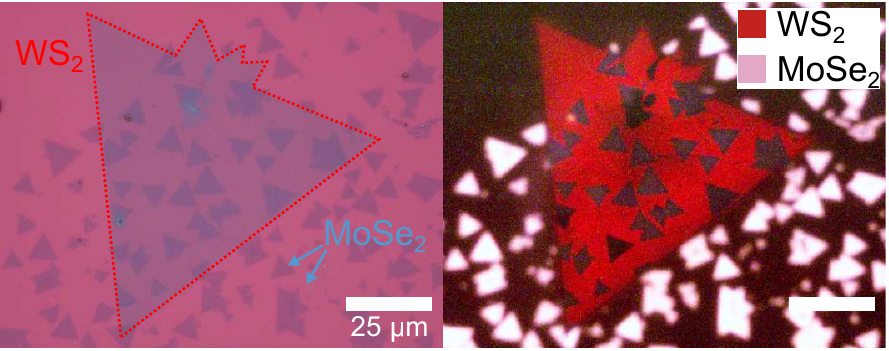}
\end{tocentry}

\begin{abstract}
Vertically stacked atomic layers from different layered crystals can be held together by van der Waals forces, which can be used for building novel heterostructures, offering a platform for developing a new generation of atomically thin, transparent and flexible devices. The performance of these devices is critically dependent on the layer thickness and the interlayer electronic coupling, influencing the hybridisation of the electronic states as well as charge and energy transfer between the layers. The electronic coupling is affected by the relative orientation of the layers as well as by the cleanliness of their interfaces. Here, we demonstrate an efficient method for monitoring interlayer coupling in heterostructures made from transition metal dichalcogenides using photoluminescence imaging in a bright-field optical microscope. The colour and brightness in such images are used here to identify mono- and few-layer crystals, and to track changes in the interlayer coupling and the emergence of interlayer excitons after thermal annealing in mechanically exfoliated flakes as well as a function of the twist angle in atomic layers grown by chemical vapour deposition. Material and crystal thickness sensitivity of the presented imaging technique makes it a powerful tool for characterisation of van der Waals heterostructures assembled by a wide variety of methods, using combinations of materials obtained through mechanical or chemical exfoliation and crystal growth.
\end{abstract}

\newpage
\subsection{Introduction}

Atomically thin materials offer a new paradigm for control of electronic excitations in the extreme two-dimensional (2D) limit in condensed matter. Recently this concept has been developed further with the creation of 2D heterostructures in which individual atomic layers are held together by van der Waals (vdW) interaction\cite{Li2015a,Geim2014,Liu2016,Novoselov}. The weak interlayer bonding loosens the lattice matching requirement, allowing a wide range of materials to be used in one device. Such vdW heterostructures combine unique properties of 2D materials with transparency and extreme flexibility, allowing a range of novel electronic and optoelectronic devices to be fabricated. Indeed, a wide variety of such devices has been demonstrated, including field-effect transistors\cite{Roy2014a,Das2014,Lee2015,Liu2015,Cui2015a,Liu2016}, light emitting devices\cite{Ross2014,Cheng2014a,Withers2015,Withers2015a}, vertical tunneling transistors\cite{Britnell2012,Britnell2013,Lin2015,Roy2015} and photodetectors\cite{Britnell2012,Buscema2015b,Buscema2014a,Furchi2014c,Lee2014,Lou2014b,Massicotte2015a,Yu2016}. 

Van der Waals heterostructures also open an attractive possibility to access interlayer excitons formed by electrons and holes localised in adjacent materials. This has recently been observed in transition metal dichalcogenide (TMD) heterobilayers with type-II band alignment\cite{Tongay2014,Fang2014,Lee2014,Lui2015,Rivera2015,Heo2015,Ceballos2015,Nayak2017}. 
Such excitons have binding energies comparable to those of their intralayer counterparts\cite{Wilson2016}, however, they can have orders of magnitude longer lifetimes due to the spatial separation of the charge carriers\cite{Rivera2015}. 
The long lifetimes in conjunction with valley-dependent optical selection rules have made interlayer excitons a promising platform for valley index manipulation and valleytronic applications\cite{Schaibley2016,Rivera2016}.

Recent advantages in growth techniques have allowed lateral\cite{Huang2014,Duan2014,Gong2015,Yoshida2015,Li2015b,Li2015c,Pant2016} and vertical\cite{Gong2014,Yu2014,Heo2015} vdW heterostructures to be manufactured by direct growth. However, the majority of heterostructures employed in research of electronic and optical properties are still created by stacking of exfoliated or chemical vapour deposition (CVD) grown crystals using polymers as transfer medium\cite{Dean2010,Castellanos-Gomez2014,Uwanno2015}. Along with the fast device prototyping, this method offers the ultimate control of the overlap and twist angle between individual layers, enabling control of the degree of the electronic coupling between them. The electronic and mechanical coupling between the layers is also affected by fabrication imperfections leading to organic residues on the crystal surfaces, which to some degree can be rectified by thermal annealing \cite{Chiu2014, Tongay2014, Yu2014, Hong2014, Zhang2016, Wang2016}. A fast method for monitoring the coupling between the layers is highly desirable, and will enable rapid assessment of the heterostructure properties and quality, that is key for fabrication of novel few atomic layer thick optoelectronic devices. 

In this paper, we present a method for rapid monitoring of the interlayer coupling in vdW heterostructures made from monolayer semiconducting TMDs using a bright-field optical microscope. We show that photoluminescence (PL) images of a large area of exfoliated or CVD-grown TMD crystals can be obtained using a standard microscope equipped with a white light source and a set of optical edge-pass filters. Using the colour and brightness of the images, the presented techniques can be utilised for rapid identification of TMD mono- and few-layers on various substrates, including polymethylmethacrylate (PMMA) and polydimethylsiloxane (PDMS) commonly used for vdW heterostructure fabrication. Furthermore, we use this method to assess the changes in the degree of the electronic coupling between the adjacent TMD crystals following thermal annealing. The microscope PL images unambiguously reveal that, while the as-fabricated TMD heterobilayers act as a set of independent monolayers because of the polymer residue between the layers, significant improvement of interlayer coupling is observed after the thermal treatment. Using PL imaging, we also investigate the coupling between individual layers in heterostructures composed of exfoliated or CVD-grown TMD monolayers with varying the interlayer twist angles. While all TMD heterobilayers show significant reduction of PL intensity due to intralayer exciton dissociation, the PL quenching is an order of magnitude stronger in samples with small rotational misalignment. The noticeable change of PL colour due to bright interlayer exciton emission can be seen in heterobilayers with aligned principal crystal axes. The sensitivity of the TMD PL emission to the individual layer thickness and coupling between different layers makes microscope PL imaging demonstrated here an indispensable tool for vdW heterostructure characterisation with wide ranging applications.

\subsection{Photoluminescence imaging of few-layer TMD samples}
\begin{figure}[ht]
	\centering
	\includegraphics{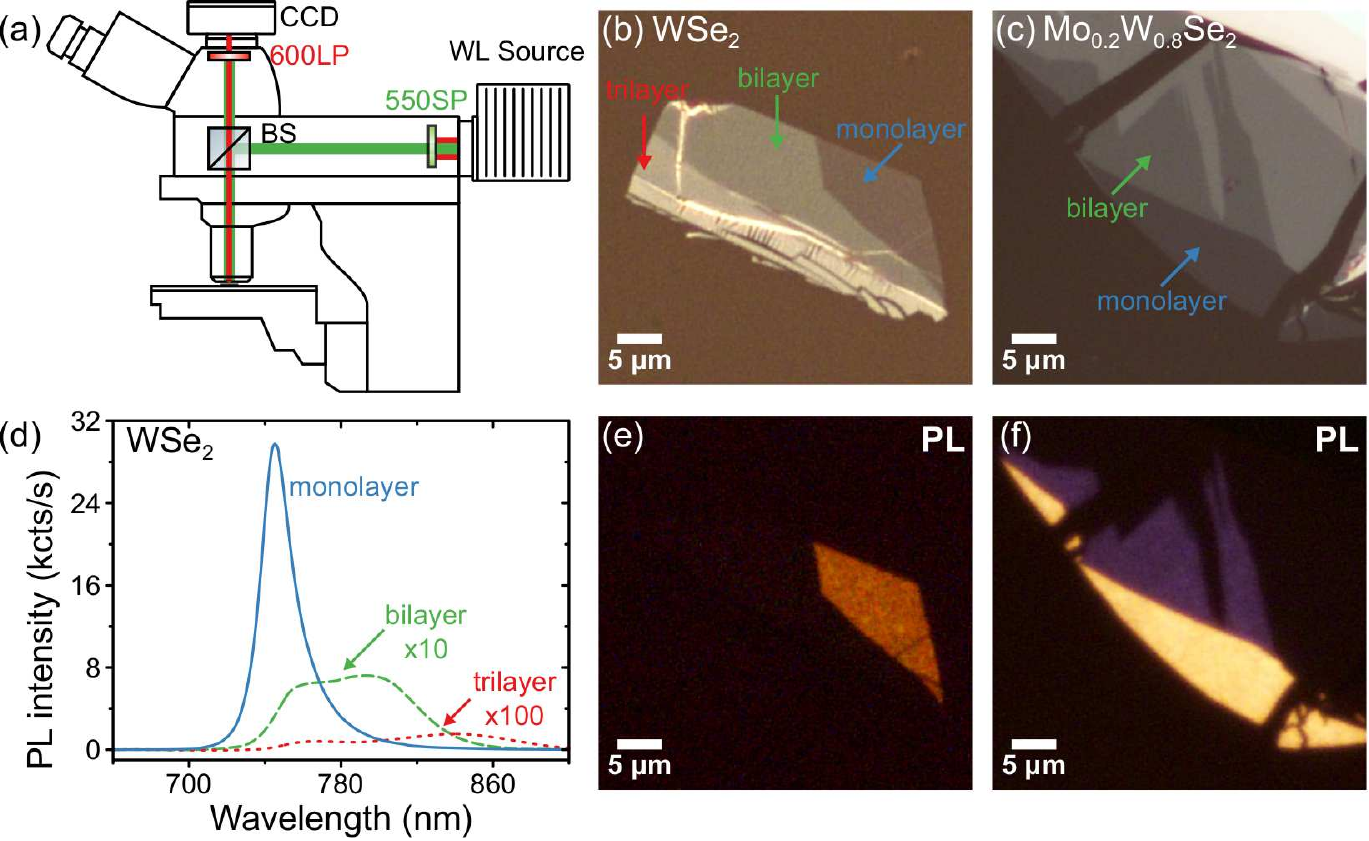}
	\caption{\textbf{Photoluminescence imaging of TMD monolayers and bilayers using a bright-field optical microscope.} (a) Schematic representation of the PL imaging setup based on the optical microscope. (b)-(c) Bright-field images of mechanically exfoliated few-atomic-layer crystals on PDMS substrates: WSe$_2$ in (b) and Mo$_{0.2}$W$_{0.8}$Se$_2$ in(c). (d) PL spectra recorded in monolayer (blue), bilayer (green) and trilayer (red) regions of the WSe$_2$ sample shown in (b). (e) PL image of the WSe$_2$ sample acquired with 1~s acquisition time and 9.6x analog gain on the camera, showing PL from the monolayer region only. (f) PL image of the Mo$_{0.2}$W$_{0.8}$Se$_2$ sample shown in (c) with clearly identifiable regions of a monolayer (yellow) and a bilayer (purple).}
	\label{fig1}
\end{figure}

The PL imaging set-up used in this study is schematically shown in Figure~\ref{fig1}~(a) and described in more detail in the Methods section below. Figure~\ref{fig1}~(b) shows a bright-field image of a WSe$_2$ flake exfoliated onto a PDMS substrate. The most translucent area in the top right corner of the flake corresponds to the monolayer region. The PL image of the same sample acquired using the experimental set-up is shown in Figure~\ref{fig1}~(e). Even with 1 second acquisition time, the monolayer region is clearly visible in the image due to bright PL emitted by the flake.
The PL emission from the WSe$_2$ bilayer region is two orders of magnitude weaker and requires longer acquisition times to be detected.

The thickness dependence of the PL intensity reflects the changes of the WSe$_2$ band structure with increasing numbers of layers\cite{Tonndorf2013}. While monolayer WSe$_2$ is a direct bandgap semiconductor, the bandgap becomes indirect for bilayers, leading to a strong quenching of the PL. Figure~\ref{fig1}~(d) compares PL spectra recorded in different regions of the flake using a separate micro-PL set up (see Methods for details). The monolayer region shows bright PL with emission peak centred at 745~nm (blue line). The direct-to-indirect bandgap transition in bilayer WSe$_2$ leads to a shift of the emission maximum to lower energies, as well as two orders of magnitude reduction of the emission intensity (dashed green line). Further increase of thickness leads to almost complete disappearance of the PL signal (dotted red line).

The abrupt change of the PL characteristics with increasing numbers of layers allows PL imaging to be used for sample thickness identification. Unlike other methods, such as optical contrast measurements\cite{Casiraghi2007}, it relies on the change of the TMD band structure and therefore its effectiveness is independent of the type of the substrate used. Figure~\ref{fig1}~(c) and (f) compares bright-field and PL images of a Mo$_{0.2}$W$_{0.8}$Se$_2$ sample exfoliated onto a PDMS substrate. Compared to the pure binary compound, the TMD alloy shows much brighter PL emission, making both monolayer and bilayer regions clearly visible in the PL image. The difference in the colour reflects the variation of emission spectrum with increasing sample thickness. Similar to WSe$_2$, the PL spectrum broadens and shifts to longer wavelengths in the bilayer regions (see Fig.~S2 in the Supplementary information). PL at longer wavelengths appears as a false purple colour in Fig.~\ref{fig1}~(f), a feature related to the transmission efficiency in the near-infrared of the colour filter arrays in the digital camera used. The thickness sensitivity of PL imaging makes it a convenient tool for rapid identification of TMD mono- and bilayers on various substrates, including PMMA and PDMS commonly used for vdW heterostructure fabrication (see more image examples in Supplementary information).

\subsection{Imaging of interlayer coupling in TMD heterobilayers}
\begin{figure}[h]
	\centering
	\includegraphics{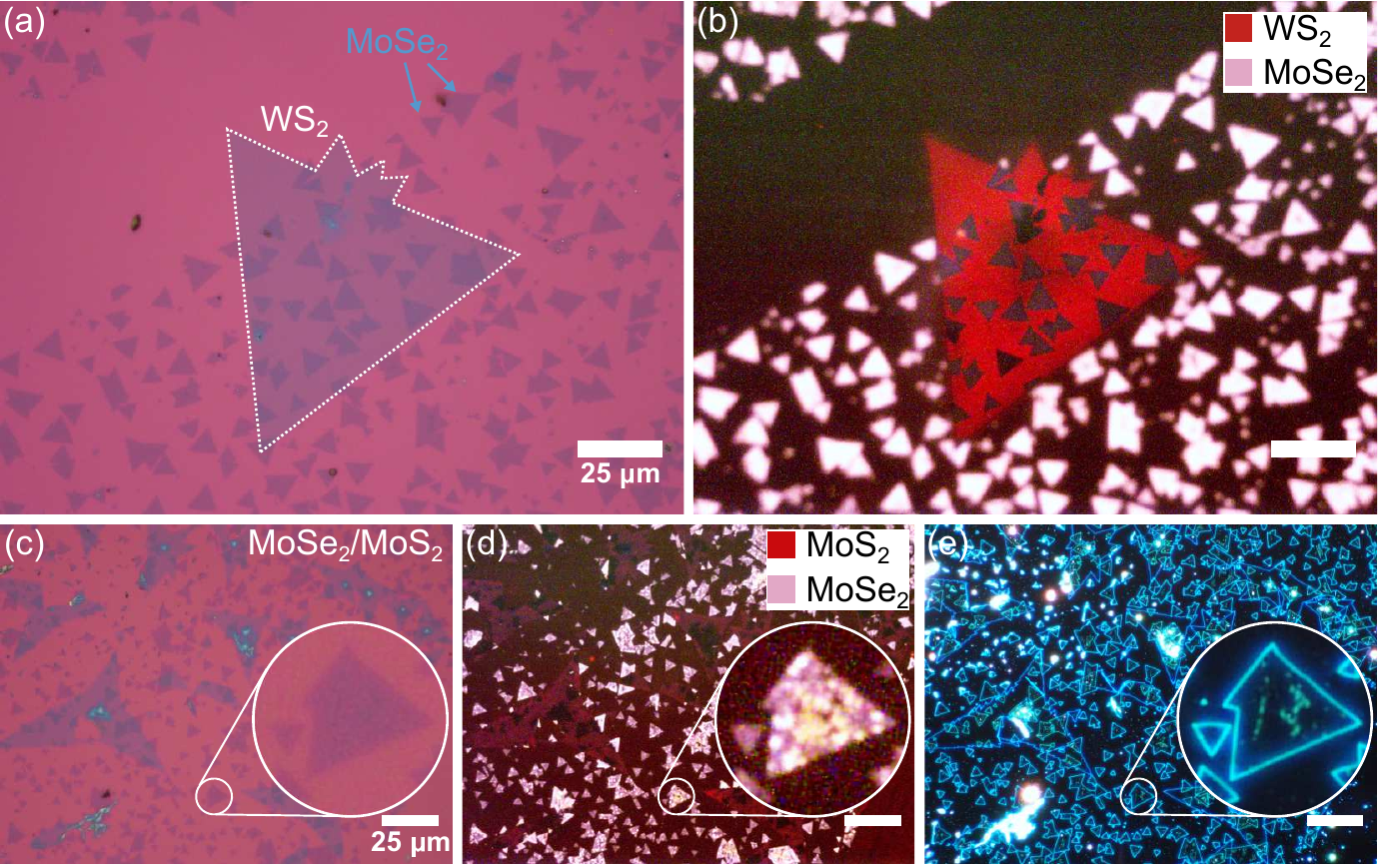}
	\caption{\textbf{Photoluminescence imaging of CVD-grown TMD samples.} (a) Bright-field image of a MoSe$_2$/WS$_2$ heterostucture on a SiO$_2$/Si substrate. (b) PL image of the sample in (a) showing that the two materials have distinctly different colours according to the wavelength of their PL bands: red for WS$_2$ and a false pink for MoSe$_2$ emitting in the near infra-red. The PL quenching in the overlap regions indicates efficient electronic coupling between two layers. (c), (d) and (e) Bright-field, PL, and dark-field images of a MoSe$_2$/MoS$_2$ heterobilayer on a SiO$_2$/Si substrate. The zoomed-in regions show the same part of the substrate in all three images, with the PL exhibiting a large amount of additional detail compared to the bright and dark microscopy images. All scale bars in the figure correspond to 25 $\mu$m.}
	\label{fig2}
\end{figure}

The wide-field nature of the PL imaging makes it especially useful for characterisation of samples produced by CVD growth, allowing large areas of a sample to be investigated at the same time. 
Figure~\ref{fig2}~(a) show a bright-field image of a MoSe$_2$/WS$_2$ sample on a SiO$_2$/Si substrate composed of CVD-grown monolayers. Large triangle in the centre of the image corresponds to single layer WS$_2$ while smaller flakes around it are MoSe$_2$ monolayers. 

Although both materials have similar appearance in the bright-field image, they can be easily distinguished by their emission colour in the PL image (Fig.~\ref{fig2}~(b)). Room-temperature PL emission of WS$_2$ is centred at 630~nm and appears dark red in the PL image, whereas MoSe$_2$ PL peaks at 790~nm and has pale pink colour. 
Here, the MoSe$_2$ monolayers have a slightly blurry appearance due to the chromatic aberration, as their PL is peaked around 800 nm. 

The bright and uniform PL in the WS$_2$ flake indicates high crystalline quality of the sample. In comparison, the CVD-grown MoS$_2$ flakes in a MoSe$_2$/MoS$_2$ sample demonstrate much weaker PL and significant variation of the emission intensity both within individual crystals and across the substrate (Fig.~\ref{fig2}~(d)). The MoSe$_2$ monolayers in the same sample also show strongly non-uniform PL emission. The observed 'grainy' structure in PL is likely caused by organic residues left from the transfer process and trapped between the flakes or the flakes and the substrate. Although the variation of emission intensity within the grainy pattern is clearly visible in the PL image, it shows little correlation with the features visible in the dark-field image (Fig.~\ref{fig2}~(e)) and is completely invisible in the bright-field image (Fig.~\ref{fig2}~(c)).

The heterobilayer regions in both samples demonstrate strong quenching of the intralayer exciton PL, indicating the efficient coupling between the layers. For the semiconducting group VI TMDs, a heterostructure formed by monolayers of two different materials will have type-II band alignment with the edges of valence and conduction bands located in different materials (further discussed below and illustrated in Fig.~\ref{fig4}~(c)). The staggered gap in TMD heterobilayers facilitates ultrafast charge separation between the two layers that acts as a dominant decay channel for optically excited intralayer excitons, significantly quenching their PL\cite{Kozawa2015,Zhang2016,Ceballos2014}.

The efficient interlayer coupling in TMD heterostructures requires the interface between adjacent layers to be clean of any contamination. As the van der Waals heterostructure fabrication through mechanical stacking relies on the use of a polymer as the transfer medium, it often results in the presence of organic residues between the atomic planes.
Figure~\ref{fig3}~(a) shows an optical image of MoSe$_2$/WSe$_2$ heterostructure mechanically assembled on a SiO$_2$/Si substrate. 
The 2D flakes exfoliated from bulk crystals onto a PDMS substrate were consecutively transferred onto the SiO$_2$/Si substrate using viscoelastic stamping\cite{Castellanos-Gomez2014} with no intermediate cleaning steps.
As it can be seen from the PL image in Fig.~\ref{fig3}~(b), the PL emitted by the heterobilayer region consists of the sum of MoSe$_2$ and WSe$_2$ emission and shows no signs of PL quenching. The unperturbed intralayer emission in the heterostructure region indicates that the interlayer coupling is suppressed by the polymer residues trapped between the layers\cite{Tongay2014, Yu2014}. 

\subsection{Tuning of interlayer coupling through thermal annealing observed in PL imaging}
\begin{figure}[h]
	\centering
	\includegraphics{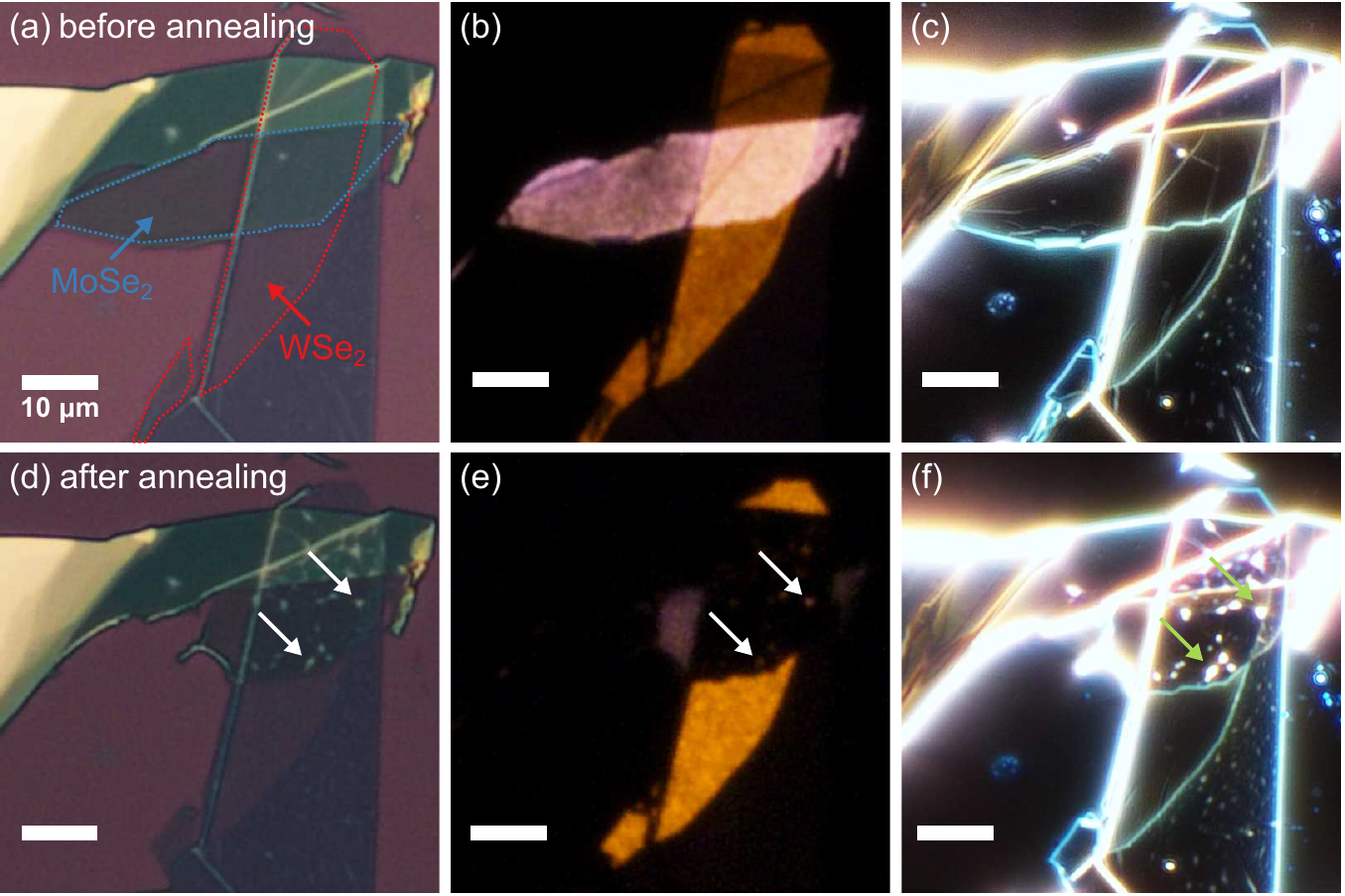}
	\caption{\textbf{Monitoring changes in the interlayer coupling introduced by thermal annealing.} (a), (b), (c) Bright-field, PL, and dark-field images of a MoSe$_2$/WSe$_2$ heterostructure fabricated using mechanical exfoliation from bulk crystals on PDMS and consequent transfer on a SiO$_2$/Si substrate. (d), (e), (f) Bright-field, PL, and dark-field images of the same structure after annealing at 120$^{\circ}$ C in high vacuum for 2 hours. Arrows show examples of contamination pockets observed in all three types of images. All scale bars in the figure correspond to 10 $\mu$m.}
	\label{fig3}
\end{figure}

Thermal annealing in vacuum or inert atmosphere is commonly used to remove organic residues from the surface of 2D crystals\cite{Chiu2014, Tongay2014, Yu2014, Hong2014, Zhang2016, Wang2016}. Here we utilise this method to improve interlayer coupling in the existing heterostructure. Figure~\ref{fig3} compares bright-field microscope images of the sample before (a) and after (d) annealing in high vacuum at 120$^{\circ}$ C for 2 hours. While a part of the isolated MoSe$_2$ monolayer was damaged during the thermal treatment, both isolated WSe$_2$ and heterostructure regions remain mostly intact.

Although the thermal treatment could not completely remove the organic residues trapped between the layers, it has caused their aggregation into small contamination pockets that can be clearly seen in both bright- and dark-field images (shown with arrows in Fig.~\ref{fig3}(d),(f)). The pockets formation is caused by the strong attraction between the two layers\cite{Vasu2016}. Comparing the dark-field images acquired before and after annealing (Fig.~\ref{fig3}~(c) and (f), respectively), it is apparent that the contamination pockets have formed only in the areas where the two crystals overlap. Moreover, we have not observed any residue aggregation in the heterostructures where the bottom crystal was annealed prior to the deposition of the top flake. The formation of contamination pockets after annealing is similar to the self-cleansing observed in vdW heterostructures\cite{Kretinin2014, Novoselov} and results in atomically clean interfaces in the contamination-free regions.

The effects of the thermal treatment can be clearly seen in the PL image of the sample in Fig.~\ref{fig3}~(e). 
Substantial increase of the emission intensity has occurred in the WSe$_2$ monolayer\cite{Wang2016, Su2016}, possibly due to removal of the polymer residues from its surface. The strongest change of the PL intensity can be seen in the layer overlap region, where both MoSe$_2$ and WSe$_2$ emission has almost completely disappeared after annealing. 
The strong quenching of the intralayer PL due to ultrafast charge separation indicates significant improvement of interlayer coupling.\cite{Rivera2015,Rivera2016}. 
While the WSe$_2$ emission intensity is significantly reduced in all areas covered by MoSe$_2$, bright PL can still be observed in the parts of the heterostructure around the contamination pockets (shown in Fig.~\ref{fig3}~(e)). The positions of these bright spots fully correlate with the pockets seen in Fig.~\ref{fig3}~(d) and (f)). However, the absence of the electronic coupling in these regions can only be revealed in the PL image in Fig.~\ref{fig3}~(e).

\subsection{Observation of formation of interlayer excitons in TMD heterobilayers using PL imaging}

The interlayer charge separation in TMD heterobilayers can lead to the formation of interlayer excitons formed by electrons and holes localised in different materials\cite{Tongay2014,Fang2014,Lui2015,Rivera2015,Heo2015,Ceballos2015,Nayak2017}. The lack of observable emission in the overlap region in Fig.~\ref{fig3}~(e) is a result of the combination of the suppression of the interlayer exciton emission in MoSe$_2$/WSe$_2$ at elevated temperatures and the relatively low efficiency of the CCD at the wavelengths above 900~nm where the PL of the interlayer exciton is expected. A further important consideration is that the momentum-space alignment of $K$ valleys in TMD heterobilayers depends of the relative layer orientation in the real space. In the sample shown in Fig.~\ref{fig3}, the crystal axes in the monolayers are not aligned, which allows spatially- and momentum-space-indirect optical transitions only, which have negligible probability. 

\begin{figure}[h]
	\centering
	\includegraphics{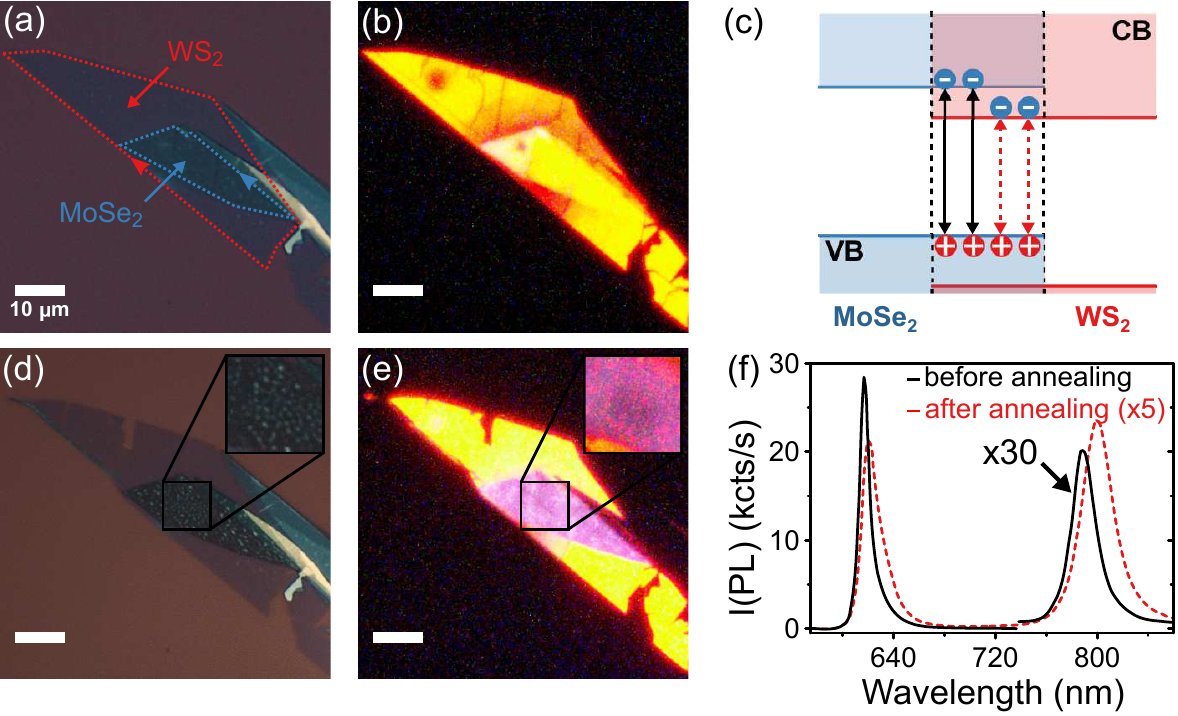}
	\caption{\textbf{Emergence of interlayer exciton PL following annealing in MoSe$_2$/WS$_2$ heterobilayers.} (a) and (b) Bright-field and PL images of a MoSe$_2$/WS$_2$ heterostructure fabricated using mechanical exfoliation from bulk crystals on PDMS and consequent transfer on a SiO$_2$/Si substrate. The crystallographic axes of the two monolayers are aligned using the the flake edges marked with arrows. (c) Band alignment of a MoSe$_2$/WS$_2$ heterostructure indicating MoSe$_2$ intralayer (solid black) and MoSe$_2$/WS$_2$ interlayer (dashed red) optical transitions. (d) and (e) Bright-field and PL images of the sample after annealing. (f) PL spectra of the heterostructure before (solid black) and after (dashed red) annealing. The intensity of the low energy peak in the spectrum measured before the annealing is multiplied by 30, whereas the intensity of the whole spectrum measured after the annealing is multiplied by 5. All scale bars in the figure correspond to 10 $\mu$m.}
	\label{fig4}
\end{figure}

In order to investigate the effects of annealing on the formation of interlayer excitons in TMD heterobilayers, we have fabricated a set of heterostructures in which crystal axes of mechanically exfoliated flakes were aligned using their terminating edges as a guide\cite{Neubeck2010}. Figure~\ref{fig4} shows the bright-field (a) and PL (b) images of a MoSe$_2$/WS$_2$ heterostructure assembled on a SiO$_2$/Si substrate using PDMS stamping. Both images were taken before annealing. The PL emission in the overlap region consists mostly of WS$_2$ PL as at room temperature it is several orders of magnitude stronger than that of MoSe$_2$\cite{Kozawa2015,Ceballos2015}. 

The improvement of the interlayer coupling leads to significant changes of the heterostructure emission that can be clearly seen in the PL image in Fig.~\ref{fig4}~(e). While isolated monolayer regions of WS$_2$ show an increase of the PL intensity due to removal of the polymer residues, the heterostructure region demonstrates a prominent change of colour, indicating a significant shift of its peak PL wavelength.

Figure~\ref{fig4}~(f) compares the PL spectra of the heterostructure before (solid black) and after (dashed red) annealing. Prior to the thermal treatment, the WS$_2$ PL is nearly two orders of magnitude stronger than that of MoSe$_2$. Unlike the isolated monolayer regions, WS$_2$ PL in the heterobilayer region is significantly quenched after annealing due to the efficient interlayer charge separation\cite{Ceballos2015,Kozawa2015}. The slight red-shift of the WS$_2$ peak is possibly a result of the change in the dielectric environment caused by reduced vertical distance between the layers \cite{Kozawa2015, Heo2015, Wang2016, Kim2016a}.

The emergence of a strong peak at 800~nm following the annealing indicates the formation of interlayer excitons. The emission energy of these excitons is defined by conduction and valence band offsets between the two materials (Fig.~\ref{fig4}~(c)). Unlike MoSe$_2$/WSe$_2$ heterobilayers, the near-degenerate conduction bands in MoSe$_2$/WS$_2$ heterostructure result in interlayer exciton states having the optical transition just a few tens of meV below the one for MoSe$_2$ \cite{Ceballos2015}. The regions of the heterostructure containing contamination pockets are visible as spots of a different colour in the PL image (magnified in Fig.~\ref{fig4}~(e)). The PL in these areas comes from both MoSe$_2$ and WS$_2$, as the aggregated residues prevent efficient coupling between two materials, causing them to act as independent layers.

\subsection{Twist angle dependence of the interlayer charge transfer}

\begin{figure}[h]
	\centering
	\includegraphics{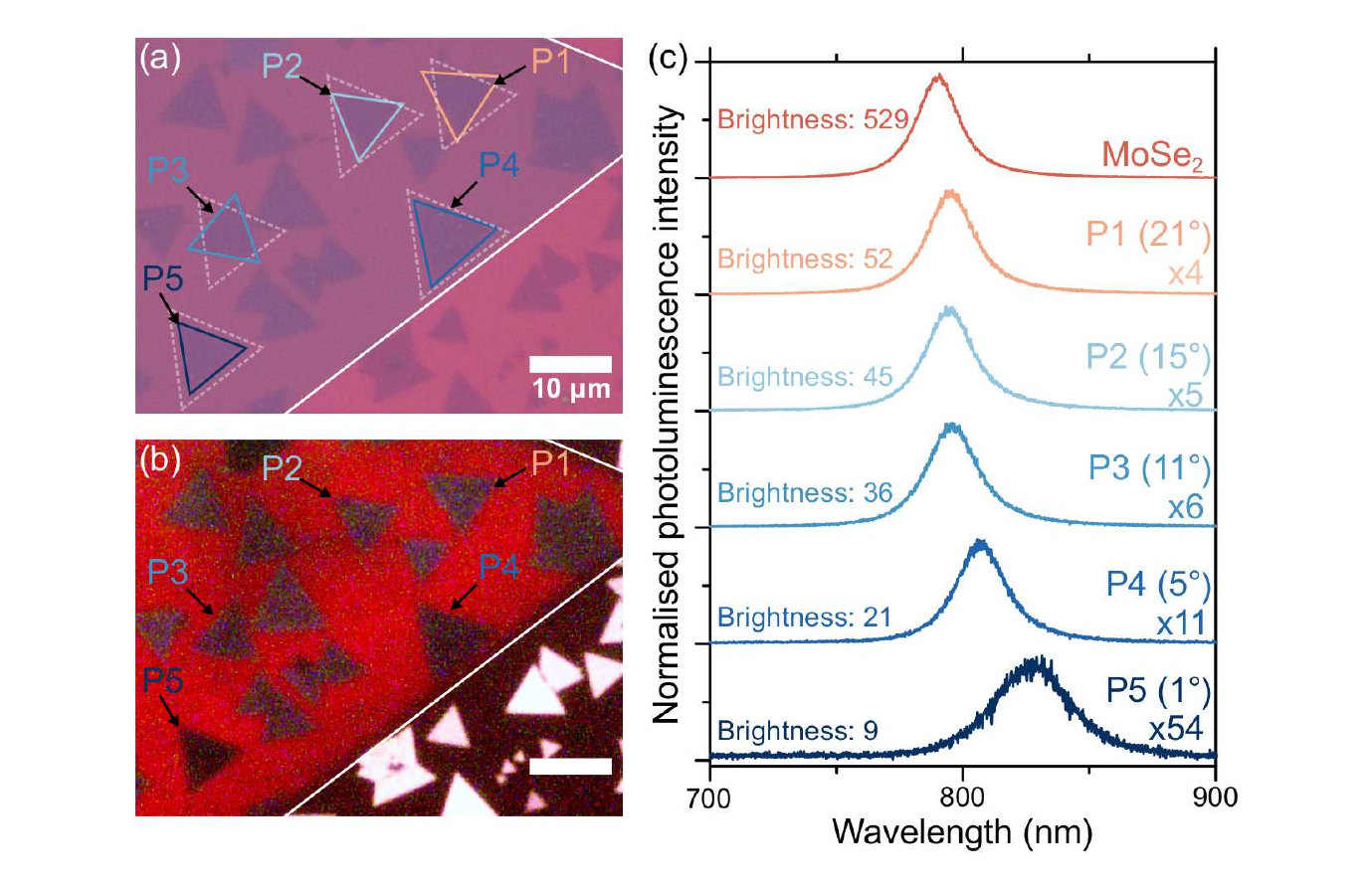}
	\caption{\textbf{Twist angle dependence of the interlayer coupling in MoSe$_2$/WS$_2$ heterobilayers.} (a) Bright-field image of a MoSe$_2$/WS$_2$ heterostructure assembled on a SiO$_2$/Si substrate from individually CVD-grown layers. Dashed white triangles indicate the orientation of the large triangular WS$_2$ flake, one visible edge of which is marked by a solid white line. In order to make the relative rotation angles more obvious, the edges of the selected MoSe$_2$ monolayers were highlighted by lines. (b) PL image of the sample showing varying degrees of the intralayer PL quenching in the overlap regions (see dark triangles). (c) PL spectra measured in the heterobilayer regions having varying twist angles (shown on the right above each spectrum). The spectra are multiplied by the factors shown on the graph (on the right above each spectrum). The brightness indicated above each curve on the left is extracted from the PL image, as explained in text.}
	\label{fig5}
\end{figure}

The PL imaging can be directly applied for studying the dependence of the interlayer coupling strength on the relative rotation between the two layers. Figure~\ref{fig5}~(a) shows a bright-field image of the MoSe$_2$/WS$_2$ sample composed of CVD-grown monolayers; dashed white line marks the edge of a large triangular WS$_2$ monolayer. Since terminating edges of triangular TMD monolayers correspond to zigzag directions\cite{Ji2015, Gong2014, Zhu2016}, the rotation angle between two layers can be easily identified by comparing the orientation of the WS$_2$ flake (indicated by dashed white triangles) with MoSe$_2$ monolayer orientation (coloured triangles). 
Figure~\ref{fig5}~(b) shows the PL image of the same region. While isolated monolayer regions of both MoSe$_2$ and WS$_2$ show bright PL, the intralayer PL intensity in both materials is significantly lowered in the overlap regions. The degree of PL quenching shows clear correlation with the interlayer twist angle with well-aligned heterobilayer regions (P4 and P5) appearing darker than regions with strong rotational misalignment (P1-P3).
Spectrally-integrated PL intensity can be extracted from the PL images by measuring the average brightness of various regions in the digital image. Here we apply this method for MoSe$_2$ triangles, both overlapping with the WS$_2$ monolayer and isolated. The average triangle image brightness is calculated as $(R+G+B)/3N$, where $N$ is the number of pixels in the triangle, and $R$, $G$ and $B$ is the intensity in the red, green and blue channels ranging between 0 and 255. The triangle brightness extracted following this procedure shows that the coherently stacked regions (twist angles of $\approx$0$^{\circ}$) have more than 5 times lower PL intensity compared to rotationally misaligned areas. 

In order to establish firmly the correlation between the PL images and detailed spectral properties of the heterobilayer regions, emission spectra were recorded in the areas with different interlayer twist angles using the micro-PL setup.
Figure~\ref{fig5}~(c) plots normalised PL spectra collected in the regions P1-P5 of the sample, as well as an isolated MoSe$_2$ monolayer. Scaling factors as well as emission intensity extracted from the PL image are listed above each curve. 
Compared to the isolated MoSe$_2$ (top curve), the heterobilayer regions demonstrate red-shifted PL with significantly lowered intensity, indicating strong interlayer coupling. It is also evident that the PL quenching becomes stronger with the decreasing interlayer twist angle. 
While the spectral position of the MoSe$_2$ intralayer exciton peak at 795~nm does not show any clear dependence on the relative orientation of the two layers, for small twist angle the emission peak red-shifts to 830~nm, indicating the change from intralayer to interlayer exciton character. The WS$_2$ PL shows strong quenching in all heterobilayer regions, however, there is no apparent correlation with the interlayer twist angle (see Fig.~S4 in the Supplementary Information).

The angular dependence of the PL intensity can be explained by the relative alignment of the MoSe$_2$ and WS$_2$ bands in the momentum space. The edges of the conduction and valence bands in TMD monolayers are located at the six $K$ points of the Brillouin zone. Originating from the in-plane orbitals of the transition atoms, these states hybridise very weakly between the layers\cite{Wilson2016}. The rotational misalignment of the two layers in the real space leads to a rotation of two Brillouin zones in momentum space. 
Therefore, the interlayer charge transfer in the vicinity of the $K$ points in the twisted heterobilayer case is a second order process, which requires phonon or defect scattering to overcome the in-plane momentum mismatch. 
As the interlayer twist angle decreases, the $K$ valleys come into alignment, significantly improving the efficiency of charge transfer between two layers. This leads to further quenching of the intralayer exciton PL. 
In coherently stacked heterobilayers, the band-gap at the Brillouin zone edge becomes direct, leading to the emergence of the interlayer exciton PL\cite{Heo2015,Nayak2017}.

\subsection{Conclusions}
In summary, we have demonstrated that rapid large-area PL imaging of 2D semiconducting TMD samples can be achieved using a standard bright-field optical microscope, rather than a dedicated optical set-up equipped with a spectrometer. The presented technique offers a highly efficient and substrate-material-independent method of flake thickness identification that can be easily combined with flake search in one experimental set-up. Furthermore, we have shown that, due to its sensitivity to interlayer charge transfer, this technique can be used to monitor the electronic coupling between individual layers in vdW heterostructures. We have successfully applied this method to investigate interlayer coupling in vdW heterostructures composed of both exfoliated and CVD-grown TMD monolayers. While the presence of organic residues between the atomic planes in TMD heterobilayers fabricated by viscoelastic stamping prevents efficient coupling between the layers, a significant improvement of the coupling efficiency and the formation of the interlayer excitons can be clearly observed in the microscope PL images of the thermally annealed samples. The presented PL imaging techniques has also been applied to assess the interlayer coupling in TMD heterobilayers having various degrees of the rotational misalignment. We have found that the degree of intralayer exciton PL quenching depends on the relative orientation of the two layers, indicating twist-angle-dependent interlayer charge transfer. The high sensitivity of the PL intensity to the charge transfer efficiency makes the presented method a very sensitive tool for investigating the coupling strength in TMD heterostructures with varying interlayer rotation and vertical separation. The short image acquisition times required makes it possible to investigate changes of the interlayer coupling in real time, allowing for example the microscope PL imaging to be used for in-situ monitoring of sample annealing or surface functionalisation. With increasing industrial and research interest in devices based on semiconductor vdW heterostructures, the PL imaging developed in this work offers a powerful characterisation method suitable for both exfoliated and CVD-grown samples at various fabrication stages.

\subsection{Methods}
\textbf{Optical microscopy system.} 
PL imaging of 2D TMD crystals was carried out using a commercial bright-field microscope (LV150N, Nikon). The schematic of the experimental setup is presented in Figure~\ref{fig1}~(a). A 550~nm short-pass filter (FESH0550, Thorlabs) was used to block the near-infrared emission from the white light source. The PL signal produced by the sample was isolated using 600~nm long-pass filter (FELH0600, Thorlabs). The short-pass (long-pass) filters were installed into the polariser (analyser) slots of the illuminator (LV-UEPI-N, Nikon), allowing quick switching between PL and bright-field imaging modes.

The PL images of the samples were acquired using a colour microscope camera (DS-Vi1, Nikon). The hot mirror mounted in front of the sensor was removed in order to enable light detection in the near-infrared range. 

\textbf{Additional micro-PL characterisation.}
Spectrally resolved PL measurements were performed in a custom-built micro-PL set-up. A diode-pumped solid state laser at 532 nm (CW532-050, Roithner) was focused onto the sample using 50x objective lens (M Plan Apo 50X, Mitutoyo). The PL signal collected in the backwards direction was isolated using a 550 nm shortpass filter (FES0550, Thorlabs) and detected by a spectrometer (SP-2-500i, Princeton Instruments) with a nitrogen cooled CCD camera (PyLoN:100BR, Princeton Instruments). All spectrally resolved PL measurements were performed at room temperature and in ambient conditions.

\textbf{Sample fabrication.}
Monolayer and few-layer TMD crystals were mechanically exfoliated from bulk crystals (provided by HQ Graphene) using wafer backgrinding tape (BT-150E-CM, Nitto). Van der Waals heterostructures were fabricated by exfoliating material onto a PDMS membrane (PF X4, Gel-Pak) followed by a transfer onto SiO$_2$/Si using a viscoelastic stamping method.

CVD WS$_2$ and MoS$_2$ crystals were grown directly on a SiO$_2$/Si substrate with a 300~nm thick SiO$_2$ layer. For heterostructure fabrication, MoSe$_2$ monolayers grown on c-plane sapphire substrates were transferred onto the substrate containing WS$_2$ and MoS$_2$ flakes using PMMA-assisted transfer - see the Supplementary Information for more details.

\begin{acknowledgement}
The authors thank the financial support of the EPSRC grant EP/M012727/1, Graphene Flagship under grant agreement 696656, and ITN Spin-NANO under grant agreement 676108. 

\end{acknowledgement}

\providecommand{\latin}[1]{#1}
\makeatletter
\providecommand{\doi}
{\begingroup\let\do\@makeother\dospecials
	\catcode`\{=1 \catcode`\}=2\doi@aux}
\providecommand{\doi@aux}[1]{\endgroup\texttt{#1}}
\makeatother
\providecommand*\mcitethebibliography{\thebibliography}
\csname @ifundefined\endcsname{endmcitethebibliography}
{\let\endmcitethebibliography\endthebibliography}{}

\end{document}